\documentclass[referee]{aa}     
\usepackage{epsf}
\usepackage{amssymb}
\usepackage{amsmath}
\usepackage{amsxtra}
\usepackage{times}
\usepackage{graphicx}

\begin{document}

\title{QSO size ratios from multiband monitoring of a microlensing high--magnification event}

\author{L. J. Goicoechea$^1$,
	V. Shalyapin$^{1,2,3}$,
	J. Gonz\'alez--Cadelo$^1$,  
	A. Oscoz$^4$}

\institute{$^1$ Departamento de F\'{\i}sica Moderna, 
	Universidad de Cantabria, 
	Avda. de Los Castros s/n, E--39005 Santander, Spain;
	\email{goicol@unican.es, juan.gonzalezc@alumnos.unican.es} \\
	$^2$ Institute of Radio Astronomy, NAS of Ukraine, 
	4 Krasnoznamennaya St., 61002, Kharkov, Ukraine \\ 
	$^3$ O. Ya. Usikov Institute of Radiophysics and Electronics, 
	NAS of Ukraine, 12 Academician Proskura St., 61085, Kharkov, 
	Ukraine; 
	\email{vshal@ire.kharkov.ua} \\
	$^4$ Instituto de Astrof\'{\i}sica de Canarias, 
	C/ V\'{\i}a L\'actea s/n, E--38200 La Laguna, Spain; 
	\email{aoscoz@ll.iac.es}}
		  
\offprints{L. J. Goicoechea}
 
\date{Submitted: January 2004}
 
\titlerunning{QSO size ratios}
\authorrunning{L. J. Goicoechea et al.}
 
\abstract{
We introduce a new scheme to study the nature of the central engine in a lensed QSO. The 
compact emission regions could have different sizes in different optical wavelengths, and our
framework permits to obtain the source size ratios when a microlensing special 
high--magnification event (e.g., a caustic crossing event, a two--dimensional maximum crossing 
event and so on) is produced in one of the QSO components. To infer the source size ratios, 
only cross--correlations between the brightness records in different optical bands are 
required. While the deconvolution method leads to a richer information (1D intrinsic luminosity
profiles), the new approach is free of the technical problems with complex inversion procedures.
Using simulations related to recent $VR$ data of Q2237+0305A, we discuss the ability of the 
scheme in the determination of the visible--to--red ratio $q = R_V/R_R$. We conclude that 
extremely accurate fluxes (with a few $\mu$Jy uncertainties, or equivalently, a few 
milli--magnitudes errors) can lead to $\sim$ 10\% measurements of $q$. Taking into account
the errors in the fluxes of Q2237+0305A from a normal ground--based telescope, $\sim$ 10  
$\mu$Jy ($\sim$ 10 mmag), it must be possible the achievement of smaller errors from the 
current superb--telescopes, and thus, an accurate determination of $q$. Obviously, to measure 
the visible--to--red ratio, the light curves cannot be contaminated by an intrinsic event or an 
important high--frequency intrinsic signal, i.e., exceeding the $\mu$Jy (mmag) level.  For an 
arbitrary lensed QSO, we finally remark that the framework seems to work better with very 
fast microlensing events.  
}

\maketitle

\section{Introduction}

In an optical component of a lensed quasar, we might see large flux variations as a result of 
gravitational microlensing. Some high--magnification events (HMEs) occur when the compact 
source crosses fold caustics. In caustic crossing events (CCEs), the folds are assumed to be 
straight lines, or more properly, the source radius and the source path are assumed to be small
compared to the caustic curvature radius. This approach is realistic for short trajectories of 
sufficiently compact sources. However, for long paths in parallel to fold caustics or 
broad--line regions crossing folds, the curvature effects can be important (e.g., Fluke \& 
Webster 1999). We also note that the actual magnification maps contain assorted caustics, so 
the curvature radii of the smallest folds are probably equal or less than the source radius, 
and obviously, our approach is not valid in this "dwarf" caustic case. The CCEs are prominent 
variations belonging to a wider family of special high-magnification events (SHMEs). The SHMEs 
are related to high--magnification regions in which the non--uniform amplification law, 
$A(X,Y)$, verifies that $A(kX,kY) = f(k) A(X,Y)$, with $k$,$f(k) >$ 0. 
 
Grieger, Kayser \& Refsdal (1988) suggested one important test on the nature
of the compact source, which is based on the analysis of the observed light curve during an 
individual CCE. They showed that the one--dimensional intrinsic luminosity profile can be 
retrieved from the brightness record of a CCE. Other physical quantities can also be determined
from observed CCEs (Grieger et al. 1988), but at present, it is not possible to fulfil all the 
observational requirements. In the nineties, the Grieger et al.'s original idea (deconvolution 
of the one--dimensional profile from an observed CCE) was developed by Grieger, Kayser \& 
Schramm (1991), Agol \& Krolik (1999), and Mineshige \& Yonehara (1999), and currently, there 
are projects to apply it. 

This paper deals with a different realistic test on the compact source structure: from the
multiband monitoring of an individual SHME, one may measure the source size ratios, which 
inform about the nature of the emitter. In the past, some authors have also done multiband 
studies of quasar microlensing (e.g., Rauch \& Blandford 1991; Wambsganss \& Paczy\'nski 
1991; Jaroszy\'nski et al. 1992; Yonehara et al. 1998; Yonehara el al. 1999). Here, we introduce 
a novel methodology to determine source
size ratios in a direct and model--independent way. A hypothetical $UBVRI$ monitoring could 
lead to a very complete set of ratios, e.g., $R_U/R_B$, $R_B/R_V$, $R_V/R_R$ and $R_R/R_I$, 
while records in only two optical bands may be used to infer one ratio. From observed events
in QSO 2237+0305 that were associated with CCEs (i.e., a kind of SHMEs), previous works 
discussed the size of the $VR$ sources (e.g., Wyithe et al. 2000; Shalyapin 2001; Yonehara 
2001, Shalyapin et al. 2002), so indirect and model--dependent estimates of $q = R_V/R_R$ are 
available for that quasar. Shalyapin et al. (2002) reported indirect measurements of $q$ for 
several circularly symmetric source models, but unfortunately, the constraints on $q$ are 
usually weak (with large uncertainties) and depend on the assumed source model. If we only 
consider the results corresponding to the best source models (power--law with the smallest 
power index and accretion disk by Shakura \& Sunyaev 1973), then it is derived a global 
interval 0.41 $\leq q \leq$ 1.26 (using 1$\sigma$ confidence limits). Therefore, for QSO 
2237+0305 and other lensed quasars, we need a new tool to obtain an accurate and robust 
estimation of $q$ or another different ratio. 

In Section 2 we present the new test. The method is robust, since it works with an arbitrary 
source model. Only similarity between the compact sources (corresponding to different optical 
filters) is required. More properly, the two--dimensional intensity distribution is arbitrary, 
but as usual, it is stationary. Therefore, while several popular scenarios are included in the 
methodology, e.g., a face--on standard disk or an inclined standard disk, we cannot discuss 
some scenarios, e.g., unstable or anisotropic (rotating) disks. In order to apply the test, we 
must focus on an observed SHME. Our method is only valid for the standard magnification close 
to a fold caustic and some non--standard amplification laws. In Section 2 we comment on a few 
non--standard behaviours leading to SHMEs. The techniques for obtaining the best value of $q$ 
as well as the criteria to measure the visible--to--red ratio are introduced in Section 3. 
Several details of the techniques are cumbersome and they are described in Appendix A. In 
Section 3, we also use synthetic light curves to test the power of the framework. The 
synthetic records are not arbitrary ones, but records related to the $V$--band and $R$--band 
GLITP (Gravitational Lensing International Time Project) microlensing peaks in the flux of 
Q2237+0305A (Alcalde et al. 2002; see also the corresponding OGLE event in Wo{\' z}niak et al. 
2000). Finally, in Section 4 we summarize and discuss our results. 

\section{The method: basic ideas}

We concentrate on a component of a multiple (gravitationally lensed) quasar. In a given 
optical band, if the mass of the lens galaxy is mainly due to main sequence stars, white 
dwarfs, black holes and so on, gravitational microlensing high--magnification fluctuations 
are expected at different epochs. The compact source travels a magnification map in the 
source plane, which contains a caustic network with folds and cusps (e.g., 
Schneider, Ehlers \& Falco 1992). In a high--magnification region, the radiation flux of the 
QSO component has two contributions: (a) a constant term ($F_0$) due to the extended source and
a possible uniform magnification of the compact source, and (b) a variable contribution caused 
by the non--uniform magnification of the compact source, which is responsible for a prominent 
event. 

We take two Cartesian coordinate frames: first, a source frame ($x$,$y$) in 
which the origin coincides with the compact source peak (the point with maximum intensity). 
The surface brightness distribution of the compact source is traced by the law $I(x,y) = I_0
B(x/R,y/R)$, where $B$ is an arbitrary function that verifies $B(0,0) = 1$ and $0 \leq B < 1$
at ($x$,$y$) $\neq$ (0,0), $R$ is the characteristic length of the intensity distribution and 
$I_0$ is the maximum intensity. Second, a magnification frame ($X$,$Y$), so the non--uniform 
magnification is $A(X,Y)$. At $t = t_0$, the magnification pattern frame coincides with the 
source frame. However, the origin of the magnification frame and the high--magnification 
region as a whole have an effective transverse motion, and we can split up the effective 
velocity in two parts: the motion parallel to the $x$-axis,  $V_{\parallel}$, and the motion 
perpendicular to that axis, $V_{\perp}$. The random stellar motions in the lens galaxy are 
implicitly neglected during the high--magnification event. At a time $t$, the global flux is 
given by
\begin{equation}
F(t) = F_0 + \frac{\epsilon I_0}{D_s^2} \int\int A[x-V_{\parallel}(t-t_0),y-V_{\perp}(t-t_0)]
B(x/R,y/R) dxdy   ,
\end{equation}
where $\epsilon$ is the dust extinction factor, which ranges from 0 (complete extinction) to 1
(no extinction), and $D_s$ is the angular diameter distance to the source. Using normalized 
coordinates $\xi = x/R$ and $\eta = y/R$, one finds 
\begin{equation}
F(t) = F_0 + \frac{\epsilon I_0 R^2}{D_s^2} \int\int 
A \{ R[\xi-V_{\parallel}(t-t_0)/R],R[\eta-V_{\perp}(t-t_0)/R] \}
B(\xi,\eta) d\xi d\eta  .
\end{equation}
On the other hand, we adopt a magnification law that is characterized by the property 
\begin{equation}
A(kX,kY) = f(k) A(X,Y) ,
\end{equation}
where $k$ is an arbitrary positive constant and $f(k)$ is another positive constant related to 
$k$. From Eqs. (2--3), a flux of the QSO component is inferred
\begin{equation}
F(t) = F_0 + \frac{\epsilon I_0 R^2 f(R)}{D_s^2} 
J\left[\frac{V_{\parallel}(t-t_0)}{R},\frac{V_{\perp}(t-t_0)}{R}\right]  .
\end{equation}
The $J$ function is given by $J[\tau,\omega] = \int\int A(\xi-\tau,\eta-\omega) B(\xi,\eta) 
d\xi d\eta$.

We remark two important issues. The first point deals with the high--magnification regions
that are consistent with the property (3), and consequently, are related to SHMEs. The 
magnification law near a cusp caustic does not verify Eq. (3) (e.g., Schneider \& Weiss 1992; 
Zakharov 1995). However, in the surroundings of a fold caustic, the behaviour of the 
non--uniform amplification is $A(X,Y) = a_C H(X)/\sqrt{X}$. Here, $H(X)$ is the Heaviside step 
function (e.g., Chang \& Refsdal 1979; Schneider \& Weiss 1987). It is evident that the 
previous standard amplification verifies Eq. (3), so the CCEs are included in our framework. 
Other non--uniform laws also agree with that equation. For example, the non--uniform 
magnifications around an one--dimensional maximum ($\alpha X^2$) and a two--dimensional maximum
($\alpha X^2 + \beta XY + \gamma Y^2$) are characterized by $f(k) = k^2$. The 
relation between the non--standard behaviours and real regions in magnification patterns merits
more attention. The source model is another important issue. With respect to this topic, we 
note that the two--dimensional intensity distribution can have circular symmetry, elliptical 
symmetry or a more complex structure, but it cannot evolve with time. 

We consider a set of compact sources that are associated with a set of optical filters, so 
all sources have the same shape and peak. They only differ in their lengths and peak 
intensities. Although this hypothesis of similarity is useful to link different sources 
corresponding to different filters, it could be false in a real situation. However, the 
similarity between sources is consistent with the standard face--on accretion disk (e.g., 
Shalyapin et al. 2002; Kochanek 2004). During a SHME, in a first optical band (number 1), the 
theoretical light curve depends on the background flux $F_{01}$, the dust extinction 
$\epsilon_1$, the maximum intensity $I_{01}$ and the length $R_1$ (see Eq. 4). In a similar 
way, in a second optical band (number 2), one has the parameters $F_{02}$, $\epsilon_2$, 
$I_{02}$ and $R_2$. We can directly compare the flux in the 1-band at the time $t$ with the 
flux in the 2-band at the time $t'$, when both times are linked from the relationship
\begin{equation} 
\frac{t-t_0}{R_1} = \frac{t'-t_0}{R_2} .
\end{equation}
Alternatively, Eq. (5) can be rewritten as
\begin{equation} 
t' = r_{21}t + (1 - r_{21})t_0  ,
\end{equation}
where $r_{21} = R_2/R_1$ is the source size ratio, and $t'$ is obtained through a dilation 
($r_{21}t$) and a delay ($t_0 - r_{21}t_0$). At the time $t'$, the 2-band flux fulfils   
\begin{equation}
\frac{D_s^2\left[F_2(t') - F_{02}\right]}{\epsilon_2 I_{02} R_2^2 f(R_2)} =
J\left[\frac{V_{\parallel}(t-t_0)}{R_1},\frac{V_{\perp}(t-t_0)}{R_1}\right]  ,
\end{equation}
so 
\begin{equation}
F_1(t) = a + bF_2(t') .
\end{equation}
The constants $a$ and $b$ are given by $a = F_{01} - F_{02}b$ and $b = \left[\epsilon_1 
I_{01} R_1^2 f(R_1)\right]/\left[\epsilon_2 I_{02} R_2^2 f(R_2)\right] >$ 0, respectively. 

Eqs. (6) and (8) show that it is viable a cross-correlation between an observed light curve 
in the first band and a brightness record in the second band, which could lead to the 
measurement of four involved parameters ($a,b,t_0,r_{21}$). We note that this new test about
the source size ratio is different to the determination of the time delay between two 
components of a lensed quasar. In the time delay estimation, using light curves in 
magnitudes, we only have two free parameters: one offset and one delay. However, in the new 
problem, there are an offset, an amplification, a characteristic epoch and a dilation factor, 
and just the dilation factor ($r_{21}$) is the relevant parameter. On the other hand, 
although the estimation of the source size ratio is possible, in practice, the observed light
curves are not continuous functions of the time and they are measured with finite accuracy. 
These observational problems (discontinuous sampling and photometric errors) may make difficult
the source size ratio estimation. Moreover, the detection of a clean SHME is not so easy, even 
if a true special event is taken place. For example, the observed event could be contaminated 
by intrinsic variability. Finally, we remember that the optical filters 1 and 2 are arbitrary 
ones, so we can apply the method to any pair of filters. From a multiband monitoring of a SHME,
one may get a very rich information. In Section 3, using simulations associated with recent 
$VR$ data of Q2237+0305A, we discuss the feasibility of accurate estimates of $q = r_{VR}$ from
the new test.

\section{Determination of the visible--to--red ratio from the method}

In the optical continuum, QSO 2237+0305 (Einstein Cross) is a gravitational mirage that 
consists of four components (A-D) round the nucleus of the deflector (lens galaxy). Irwin et 
al. (1989) discovered microlensing variability in that system, and after such a fascinating 
discovery, several groups did an important effort to monitor its components (e.g., Corrigan et 
al. 1991; {\O}stensen et al. 1996; Vakulik et al. 1997; Wo{\' z}niak et al. 2000; Alcalde et 
al. 2002; Schmidt et al. 2002). In recent dates, the OGLE collaboration presented the first 
detailed light curves of Q2237+0305A--D in the $V$ band, which showed two clear 
high--magnification events between days (in JD--2450000) 1200 and 1800: one in the A component 
and another one in the C component (Wo{\' z}niak et al. 2000). The GLITP collaboration also 
reported excellent $V$--band and $R$--band records of the four QSO components, which are 
complementary to the OGLE data (Alcalde et al. 2002). The GLITP monitoring (from day 1450 to 
day 1575) permitted to accurately trace the behaviour around the maximum of the fluctuation in 
the A component, so that the OGLE--GLITP event in Q2237+0305A is by far the best observed 
high--magnification variation in the Einstein Cross. We call OGLE--GLITP/Q2237+0305A event to 
the peak between days 1400 and 1600. 

If we have $VR$ observations of a SHME in a QSO component, we may robustly measure the 
parameter $q = R_V/R_R$. Eqs. (6) and (8) are rewritten as $t' = qt + (1 - q)t_0$ and $F_R(t) =
a + bF_V(t')$, respectively, and it might be possible to infer the 
visible--to--red ratio from a comparison between the $V$--band record and the $R$--band one. 
While the measurement would be robust because it would not depend on particular (stationary)
source models, the discontinuous sampling and the photometric uncertainties could be obstacles 
to get an estimation of $q$. First, we present some techniques for obtaining the best value of 
$q$. The criteria to measure the ratio are also quoted. Second, in order to test the power of 
the framework, it is applied to synthetic light curves (simulations) related to the 
GLITP/Q2237+0305A data. The influence of several observational parameters is analyzed in detail.

\subsection{Best value and measurement of the source size ratio $R_V/R_R$}

As a first method to infer the best value of $q$ (or equivalently, the best ratio), we use a 
usual $\chi^2$ minimization. As a second estimator, we use the dispersion, which is also 
popular between people working on 
discrete time series. Thinking of the measurement of time delays, Pelt and collaborators (Pelt 
et al. 1994, 1996) developed this statistical technique. During the last decade, the $D^2$ 
minimization was successfully applied in the time delay estimation of several lens systems, so 
the method is able to find delays in gravitational mirages. Unfortunately, the original version
of the estimator is not useful in our problem (see comments in the last paragraph of section 2),
and therefore, we must slightly modify the original scheme by Pelt et al. Finally, we take a 
variant of the minimum dispersion method, which is called $\epsilon^2$ minimization. All the 
three techniques are described in Appendix A. In the Appendix we comment on the main details of
the minimization techniques, including the way to transform the four dimensional estimators 
into three dimensional ($\chi^2$) or two dimensional ($D^2$ and $\epsilon^2$) ones. 

For a given estimator of $q$, we follow two different approaches. In the first framework, we 
make one {\it repetition} of the experiment by adding a random quantity to each original flux 
in the light curves. The random quantities are realizations of normal distributions around 
zero, with standard deviations equal to the errors of the fluxes. We can make a large number of
{\it repetitions}, and thus, obtain a large number of $q$ values. The true value will be 
included in the whole distribution of {\it measured} ratios. This first procedure is called 
NORMAL. In the second approach, we use a bootstrap procedure (BOOTSTRAP). The {\it repetitions}
are generated by bootstrap resampling of residuals from the original light curves smoothed by a
filter. In section 3.3 (see below), each original light curve is smoothed by a minimum 
(3--point) filter, and the results seem to be stable with respect to the choice of any 
reasonable smoothing filter. The smoothed curves are assumed to be rough reconstructions of the
underlying signals, and thus, the residuals are taken as errors, which may be resampled to 
infer a bootstrap simulation. Each bootstrap simulation of both the $V$--band and 
$R$--band records is considered as one {\it repetition} of the original experiment. All our $q$ 
distributions include 300 ratios. For each distribution, we study the feasibility of an 
accurate determination of $q$. The source size ratio is only measured when the true value of 
$q$ is included in the $q$ range for the dominant feature. In that case, after the cleaning of 
the distribution, we compute the centre of the main structure (central value) and the standard 
error (about 70\% confidence interval). To measure the ratio, we properly clean the 
distribution of values of $q$, i.e., it is dropped the signal out of the dominant structure and
its wings. 

In order to clarify the difference between some expressions, we comment them all together. 
First, the term "true value" refers to the true value of a physical parameter. In a real 
experiment, the true value is unknown and we want to estimate it. However, in a synthetic 
experiment, the astronomer controls the involved physics and chooses the true value. Second, 
the term "best value" (or "best solution") refers to the direct estimation of a physical 
quantity from an experiment and a technique ($\chi^2$ minimization, minimum dispersion method 
or minimum modified dispersion method). Third, after the direct estimation, we make
{\it repetitions} (NORMAL or BOOTSTRAP) of the experiment, obtain a best value from each 
{\it repetition}, and study the distribution of best values. The distribution may include
"peaks" for some values of the physical parameter, and sometimes a "dominant peak" may appear. 
In a favourable situation, the true value, the best value and the central value for the 
dominant structure would be nearby each others.      

\begin{figure}[hbtp]
\centering
\epsfxsize=10.00 cm
\rotatebox{0}{\epsffile{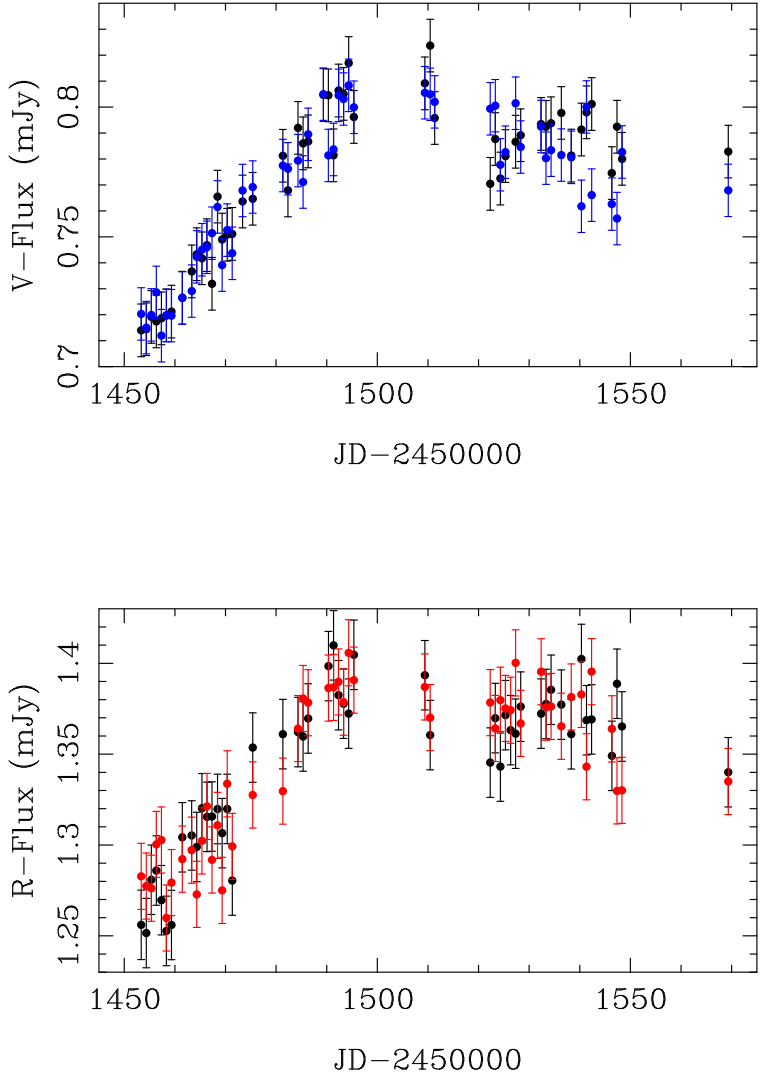}}
\caption[]{GLITP (black symbols) and SYNa (blue and red symbols) $VR$ light curves. The GLITP
fluxes correspond to observations of Q2237+0305A, whereas the SYNa fluxes are synthetic data. 
In the simulations (SYNa), we take noise processes and sampling consistent with the GLITP
photometric errors and dates.}
\end{figure}

\begin{figure}[hbtp]
\centering
\epsfxsize=10.00 cm
\rotatebox{0}{\epsffile{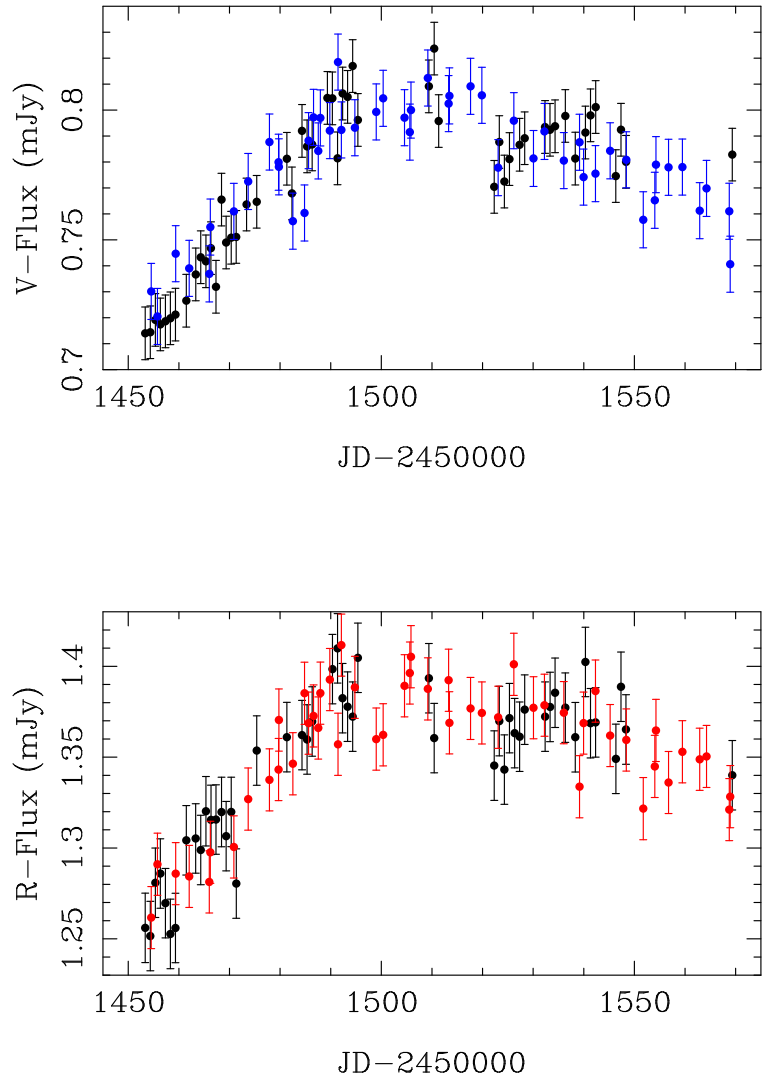}}
\caption[]{GLITP (black symbols) and SYNd (blue and red symbols) $VR$ light curves. The simulations 
(SYNd) have a homogeneous sampling.}
\end{figure}

\begin{figure}[hbtp]
\centering
\epsfxsize=10.00 cm
\rotatebox{0}{\epsffile{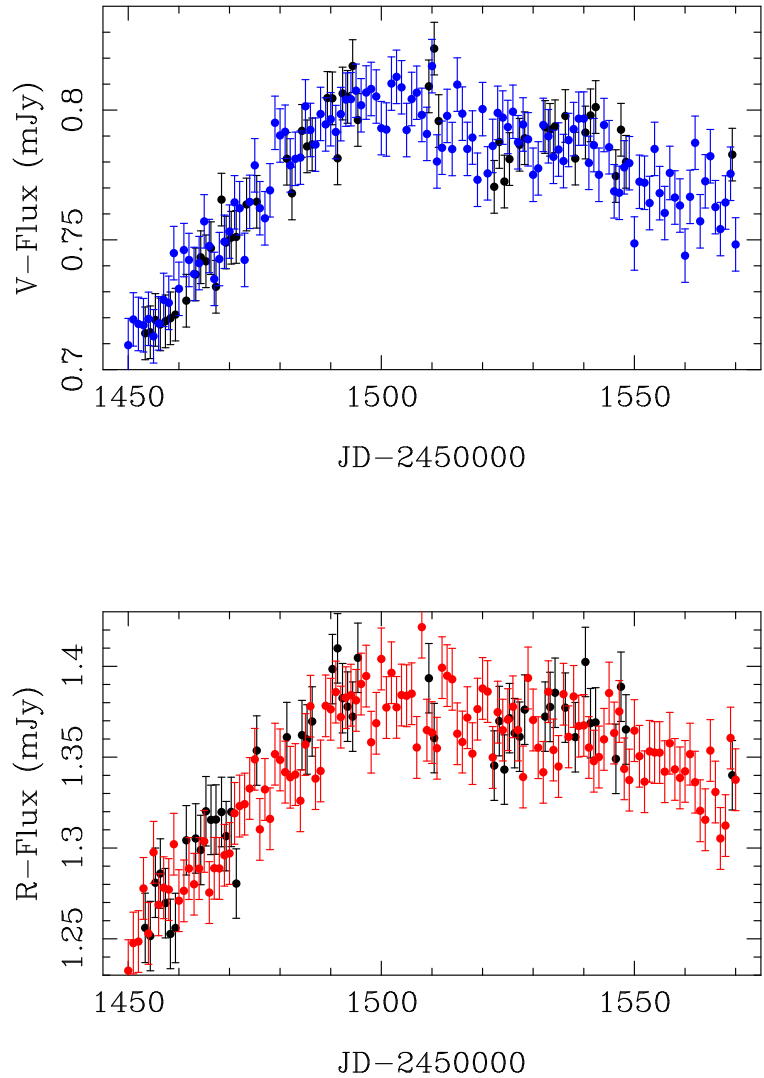}}
\caption[]{GLITP (black symbols) and SYNe (blue and red symbols) $VR$ light curves. The new 
synthetic records (SYNe) have a quasi--continuous sampling of one {\it photometric 
measurement} per day.}
\end{figure}

\begin{figure}[hbtp]
\centering
\epsfxsize=10.00 cm
\rotatebox{0}{\epsffile{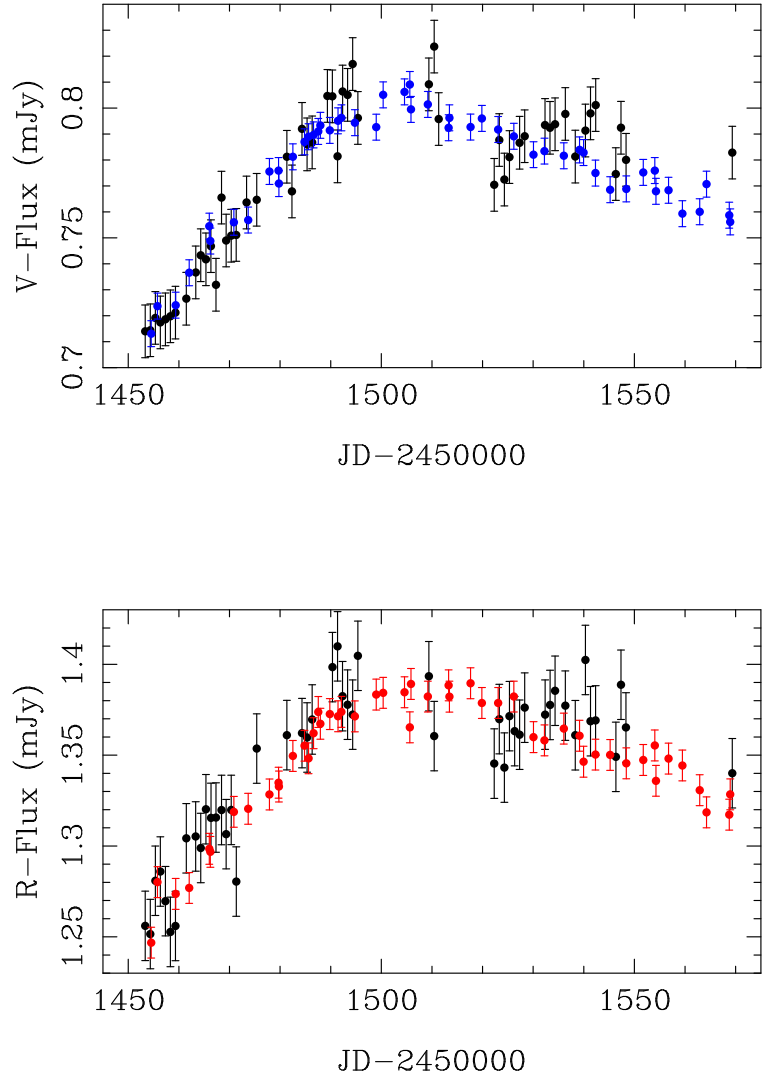}}
\caption[]{GLITP (black symbols) and SYNf (blue and red symbols) data sets. The simulations (SYNf)
have a homogeneous sampling and flux errors two times less than the GLITP ones.}
\end{figure}

\begin{figure}[hbtp]
\centering
\epsfxsize=10.00 cm
\rotatebox{0}{\epsffile{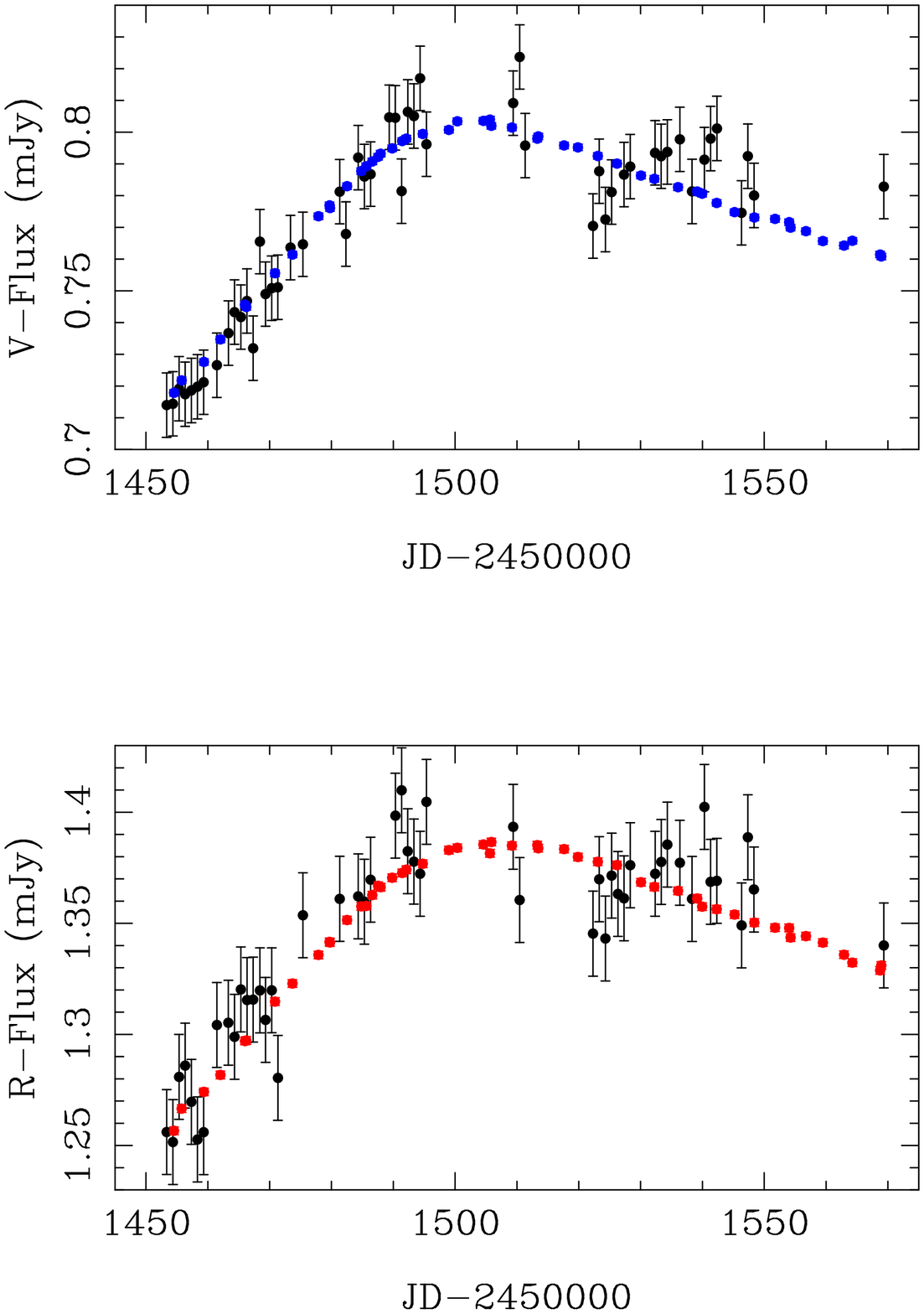}}
\caption[]{GLITP (black symbols) and SYNg (blue and red symbols) data sets. The
simulations (SYNg) have a homogeneous sampling and uncertainties 10 times less than the GLITP errors.}
\end{figure}

\subsection{Synthetic light curves}

To study the ability of our scheme, we make synthetic $VR$ light curves and apply the formalism
to the simulations (see section 3.3). The GLITP observations of Q2237+0305A are used as a 
reference data set, which contains 49 fluxes in the $R$ band and 52 fluxes in the $V$ band. The
observations were made with the 2.56 m Nordic Optical Telescope (NOT) at Canary Islands (Spain).
By pure chance, the component was monitored from day 1450 to day 1575 (in JD--2450000), just 
during a microlensing peak (Alcalde et al. 2002). In each optical filter, to compute a typical 
error in the fluxes, the GLITP collaboration used the mean of the absolute differences between 
adjacent days. Errors of $\sigma_R$ = 0.017 mJy and $\sigma_V$ = 0.010 mJy were derived from 
that procedure. We note that our $R$--band calibration is made in an arbitrary way, so our 
$R$--band fluxes disagree with those in the GLITP Web site. However, the inappropriate 
calibration is not a problem, because the $q$ estimates do not depend on the calibration of the
$VR$ light curves. The amplitudes of the observed features are about 10 times the photometric 
uncertainties.

Going into details, each underlying signal is generated through the Eq. (13) in Shalyapin et 
al. (2002). Therefore, a circularly symmetric source model and a caustic crossing are involved.
We use a $p$ = 3/2 power--law intensity profile. The source size ratio and the time of caustic 
crossing by the common center of the sources are taken as $q$ = 0.8 and $t_0$ = 1483, 
respectively. Moreover, the final fluxes are normally distributed around the underlying ones, 
so the {\it observational} random noise is characterized by a normal standard deviation. Except
for a family of experiments (those incorporating extremely accurate fluxes), the error bar is 
computed from the GLITP criterion (the mean of the absolute differences between adjacent 
fluxes), and it is close to the normal standard deviation. All the simulated $VR$ records have 
time coverage and flux variations consistent with the GLITP $VR$ records for Q2237+0305A. In 
the first synthetic experiments, the flux errors and sampling properties agree with the GLITP 
photometric uncertainties and sampling. The sampling properties are modified in a second kind 
of experiments, whereas homogeneously distributed and very accurate fluxes are produced in the 
last experiments. 

Firstly, it is considered the SYNa data set. In Figure 1 we have drawn together the observed 
(GLITP) light curves and the SYNa ones. The black symbols represent the observations, while 
the blue and red symbols represent the simulations. In SYNa, the standard deviations of the 
{\it observational} noise processes are assumed to be 0.017 mJy ($R$--band) and 0.010 mJy 
($V$--band), i.e., in agreement with the GLITP uncertainties. The SYNa sampling also 
coincides with the GLITP distribution of dates. Indeed, in Fig. 1 we can see synthetic data 
very similar to the GLITP ones. Besides of this first GLITP--like data set, we generate two 
more GLITP--like experiments. They are called SYNb and SYNc simulations, and the 
observations and simulations are again similar in all the details.

An interesting issue is the influence of some observational aspects on the determination of 
$q$, e.g., the sampling. Therefore, in a second kind of simulations, we exclusively modify the
sampling properties. The sampling of the light curves has two main properties: rate and 
homogeneity. From a ground--based telescope, a sampling rate of one frame each two or three 
days is excellent. This very good rate is used to generate the first kind of simulations 
(GLITP--like simulations). The sampling quality also depends on the homogeneity in the dates. 
For example, the GLITP--like light curves have important gaps. In the SYNd experiment (see 
Figure 2), we only improve the sampling homogeneity. There are 50 fluxes in each optical band 
(blue and red symbols in Fig. 2), but their time distributions are highly homogeneous. In 
another experiment (SYNe) we consider a quasi--continuous sampling of one {\it photometric 
measurement} per day. The GLITP records (black symbols) and the SYNe light curves (blue and 
red symbols) are showed in Figure 3.

From a third kind of experiments, we can also test the influence of the flux uncertainties. 
The SYNf $VR$ light curves are generated with normal standard deviations ({\it observational}
noise processes) of 0.008 mJy ($R$--band) and 0.005 mJy ($V$--band). To produce these synthetic
light curves, the sampling rate is assumed to be the GLITP one, the dates are homogeneously 
distributed along the time coverage, and the flux errors are lowered by a factor of 2. In 
Figure 4, the black symbols represent the GLITP data, and the blue and red symbols represent 
the SYNf data. Both data sets (observations and simulations) are similar in some details. 
However, apart from the sampling homogeneity, it is clear in Fig. 4 that the simulated errors 
are half the GLITP ones. The SYNg light curves are characterized by normal standard 
deviations of 1.7 $\mu$Jy ($R$--band) and 1 $\mu$Jy ($V$--band), i.e., with respect to the 
GLITP--like curves, the uncertainties are lowered by a factor of 10. In Figure 5, we show the 
two synthetic curves. The accuracy is impressive, and each error bar has a size similar to the 
point size. In the SYNg experiment, the error bars are not inferred from the usual criterion
(the mean of the absolute differences between adjacent fluxes), but they are chosen to be the 
normal standard deviations. The hypothetical observer cannot measure a flux uncertainty (in a 
given optical band) from the scatter of fluxes, because the error is clearly smaller than the 
true day--to--day variability. We assume that the observer uses a non--biased criterion. 
Moreover, we also use 10 additional experiments similar to SYNf and 10 additional data sets
similar to SYNg. 

Although our study is mainly based on 27 data sets, we have produced and analyzed about 100 
synthetic experiments ("SYN"). The names and properties of the seven basic experiments (see 
here above) are listed in Table 1. 

\begin{table}
\centering
\begin{tabular}{ccccc}
\hline\noalign{\smallskip}
Name & Sampling rate (data/week) & Homogeneity & $V$--band noise ($\mu$Jy) & 
$R$--band noise ($\mu$Jy) \\
\noalign{\smallskip}\hline\noalign{\smallskip}
SYNa  & $\sim$ 3 & No  & 10 & 17  \\
SYNb  & $\sim$ 3 & No  & 10 & 17  \\
SYNc  & $\sim$ 3 & No  & 10 & 17  \\
SYNd  & $\sim$ 3 & Yes & 10 & 17  \\
SYNe  & 7        & Yes & 10 & 17  \\
SYNf  & $\sim$ 3 & Yes &  5 &  8  \\
SYNg  & $\sim$ 3 & Yes &  1 & 1.7 \\
\noalign{\smallskip}\hline
\end{tabular}
\caption{Basic synthetic experiments.\label{tbl-1}}
\end{table}

\begin{figure}[hbtp]
\centering
\epsfxsize=10.00 cm
\rotatebox{0}{\epsffile{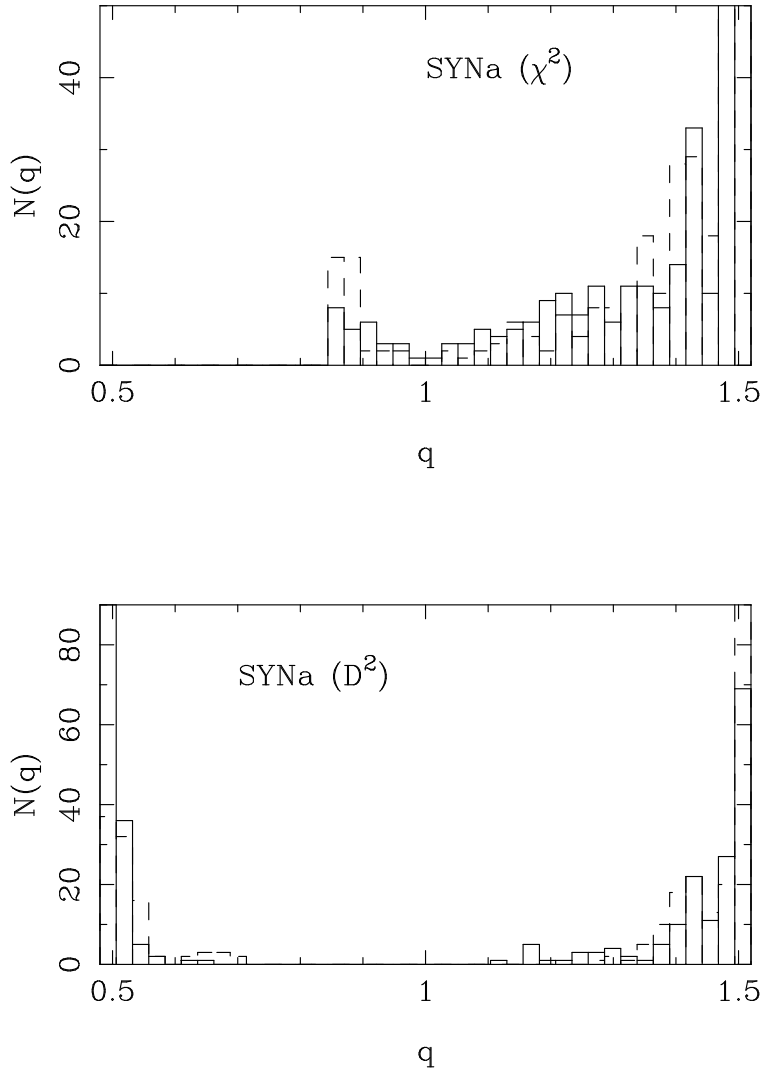}}
\caption[]{Distributions of $q$ based on NORMAL and BOOTSTRAP repetitions of the SYNa 
experiment. To derive the BOOTSTRAP repetitions, we use curves smoothed from a 3--point filter
(a time window of about 5 days). Dashed lines represent the NORMAL distributions and solid
lines trace the BOOTSTRAP histograms. {\it Top panel}: minimum $\chi^2$ method ($\alpha$ = 2.5 
days). {\it Bottom panel}: minimum dispersion method ($\delta$ = 2.5 days).}
\end{figure}

\begin{figure}[hbtp]
\centering
\epsfxsize=10.00 cm
\rotatebox{0}{\epsffile{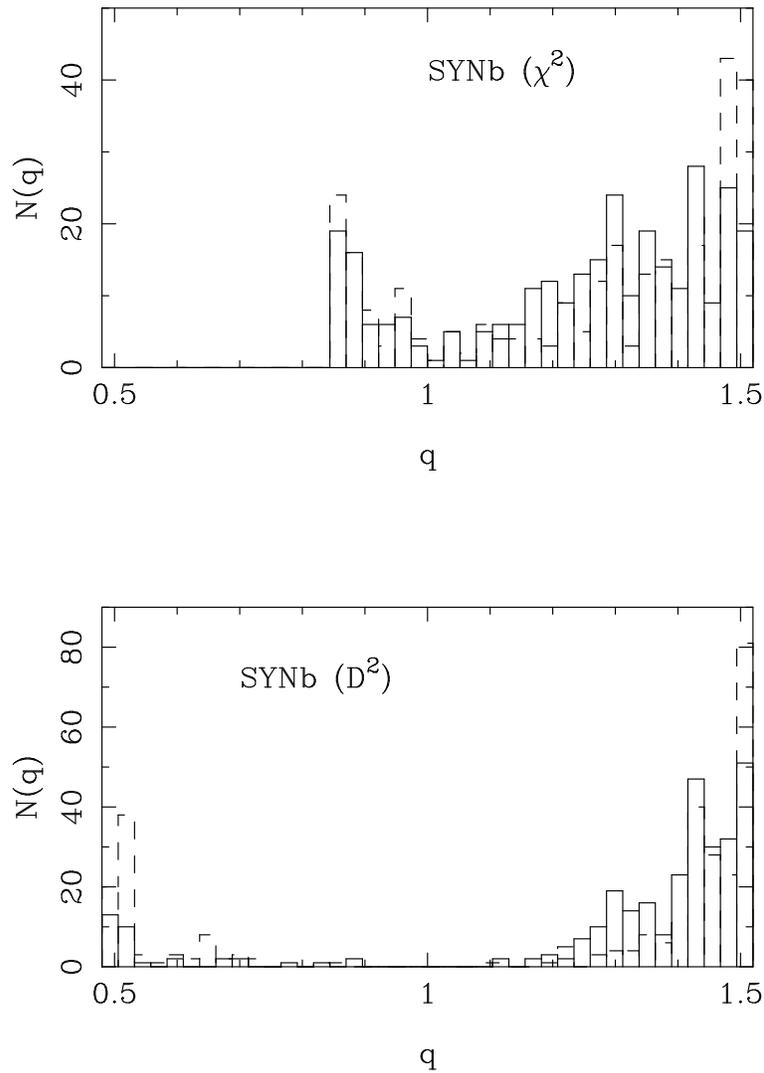}}
\caption[]{Distributions of $q$ from NORMAL and BOOTSTRAP repetitions of the SYNb experiment
(GLITP--like simulations). In the BOOTSTRAP procedure, it is used a 3--point filter (a time 
window of about 5 days). As remarked in the caption under Fig. 2, dashed lines and solid lines 
represent the NORMAL and BOOTSTRAP distributions, respectively. {\it Top panel}: $\chi^2$ 
($\alpha$ = 2.5 days). {\it Bottom panel}: $D^2$ ($\delta$ = 2.5 days).}
\end{figure}

\begin{figure}[hbtp]
\centering
\epsfxsize=10.00 cm
\rotatebox{0}{\epsffile{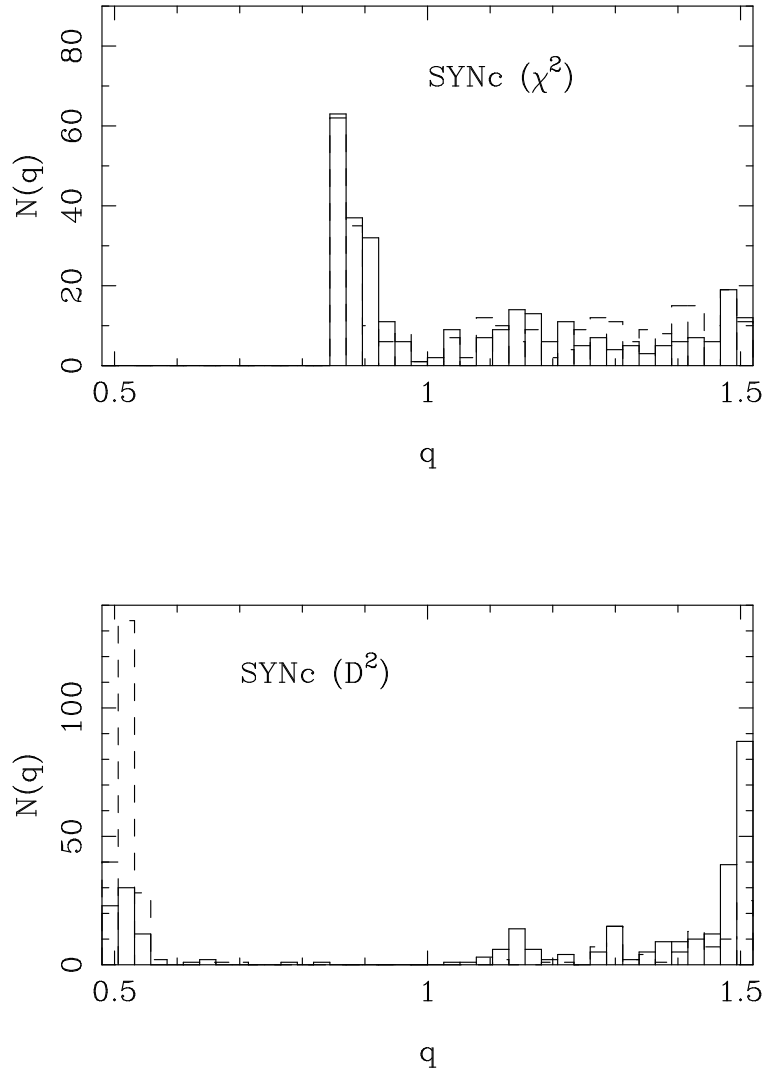}}
\caption[]{Histograms based on repetitions of the SYNc experiment (GLITP--like simulations). 
We present NORMAL (dashed lines) and BOOTSTRAP (solid lines) distributions. We use a 3--point 
filter in the BOOTSTRAP scheme. {\it Top panel}: $\chi^2$ ($\alpha$ = 2.5 days). {\it Bottom 
panel}: $D^2$ ($\delta$ = 2.5 days).}
\end{figure}

\begin{figure}[hbtp]
\centering
\epsfxsize=10.00 cm
\rotatebox{0}{\epsffile{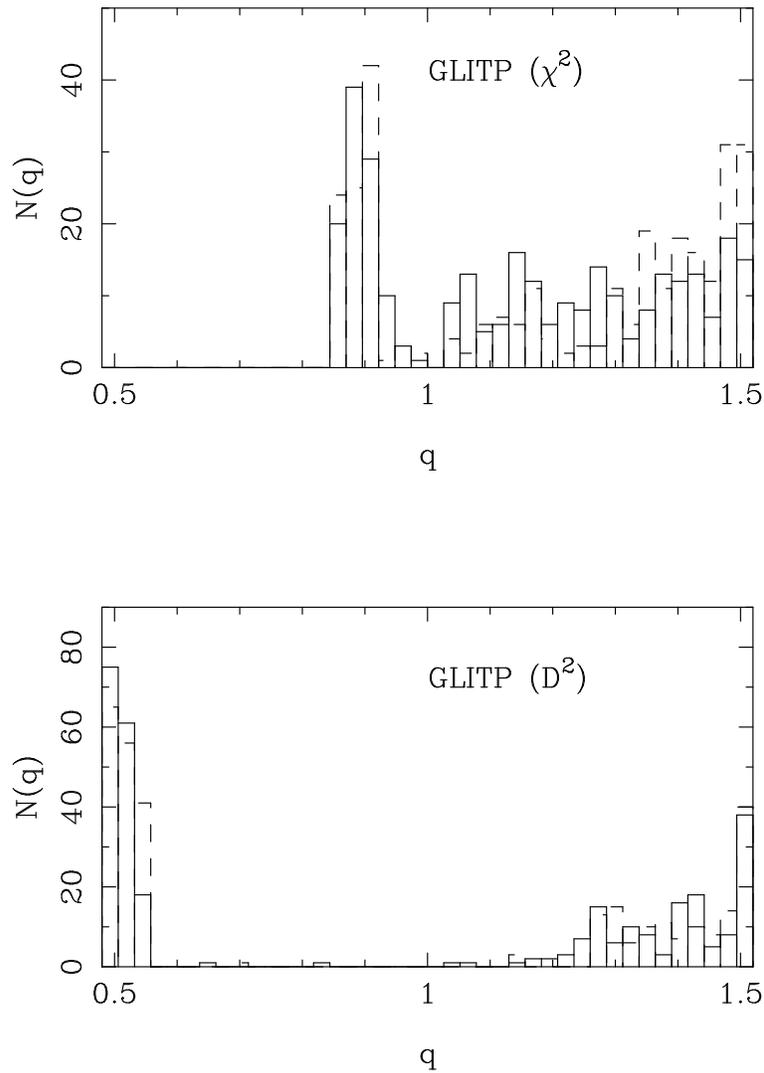}}
\caption[]{NORMAL (dashed lines) and BOOTSTRAP (solid lines) histograms associated with 
repetitions of the GLITP experiment. We show the results from a study with high time 
resolution: $\alpha$ = $\delta$ = 2.5 days and a 3--point filter.}
\end{figure}

\begin{figure}[hbtp]
\centering
\epsfxsize=10.00 cm
\rotatebox{0}{\epsffile{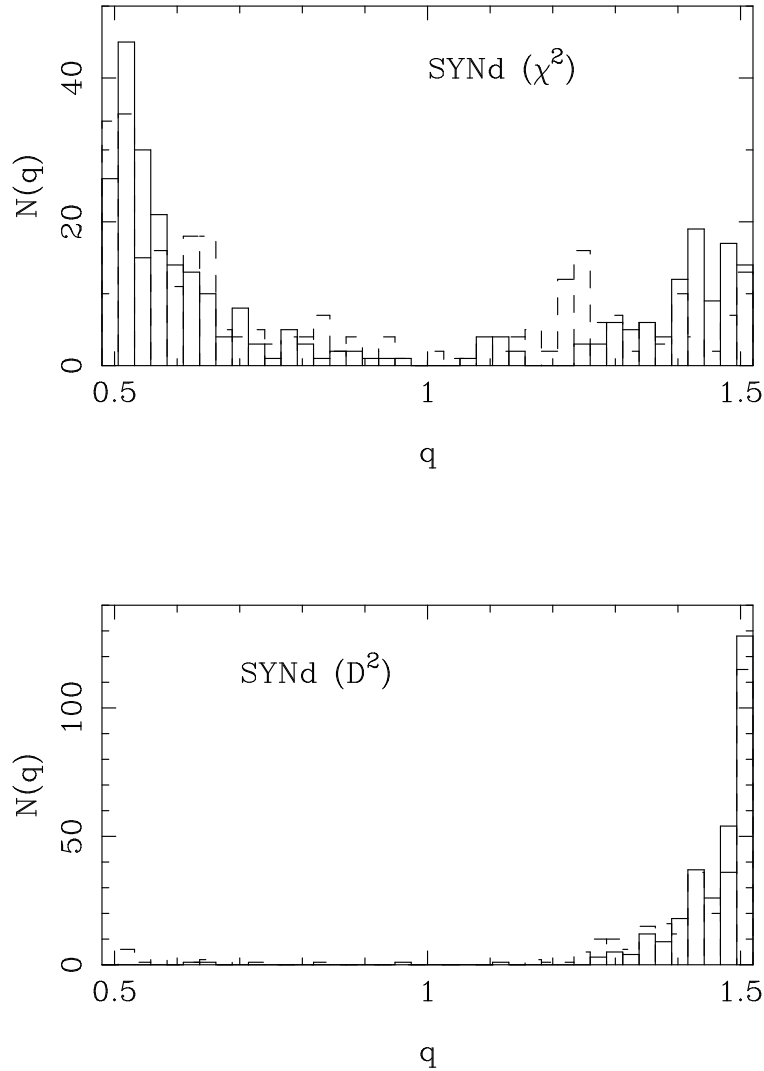}}
\caption[]{NORMAL (dashed lines) and BOOTSTRAP (solid lines) histograms associated with 
repetitions of the SYNd experiment. We show the results from the analysis with $\alpha$ = 
$\delta$ = 2.5 days and a 3--point filter.}
\end{figure}

\begin{figure}[hbtp]
\centering
\epsfxsize=10.00 cm
\rotatebox{0}{\epsffile{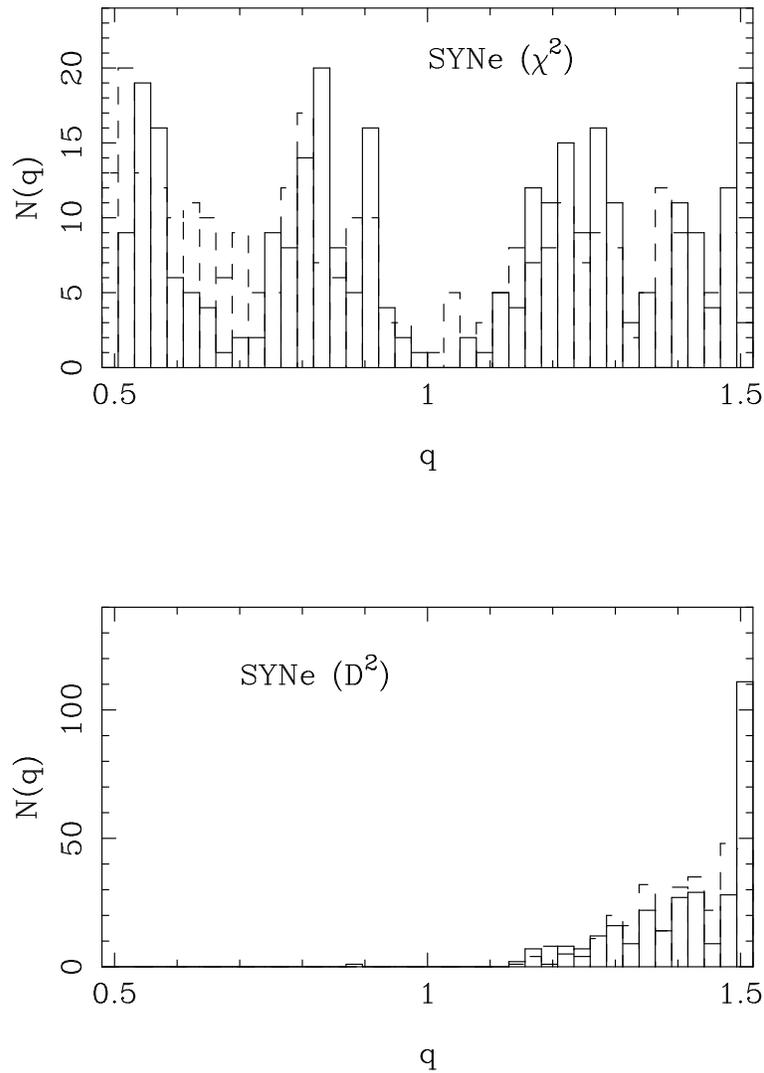}}
\caption[]{NORMAL (dashed lines) and BOOTSTRAP (solid lines) histograms associated with 
repetitions of the SYNe experiment. We work with very high time resolution: $\alpha$ = 
$\delta$ = 1.2 days and a 3--point filter.}
\end{figure}

\begin{figure}[hbtp]
\centering
\epsfxsize=10.00 cm
\rotatebox{0}{\epsffile{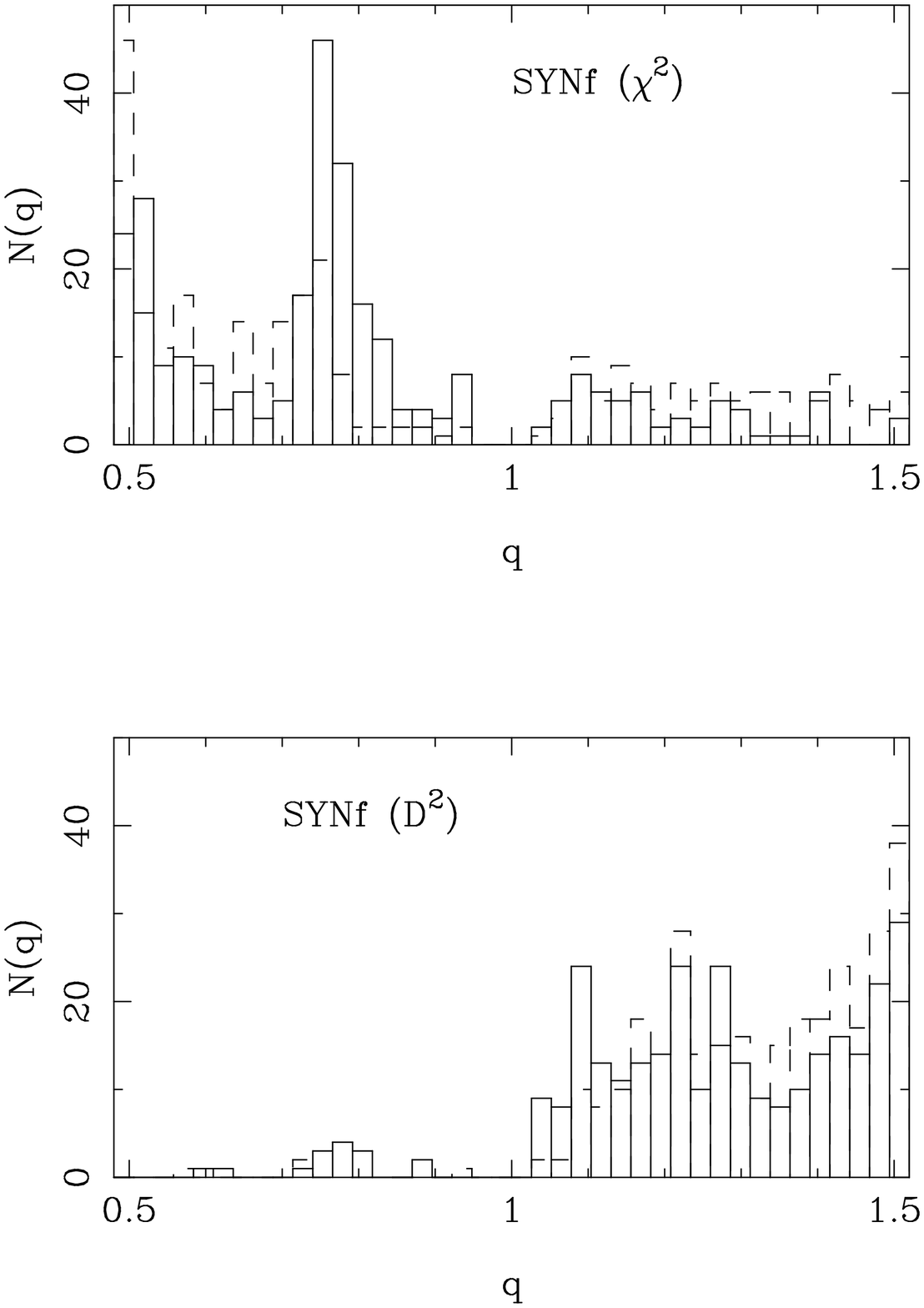}}
\caption[]{NORMAL (dashed lines) and BOOTSTRAP (solid lines) histograms associated with 
repetitions of the SYNf experiment.}
\end{figure}

\begin{figure}[hbtp]
\centering
\epsfxsize=10.00 cm
\rotatebox{0}{\epsffile{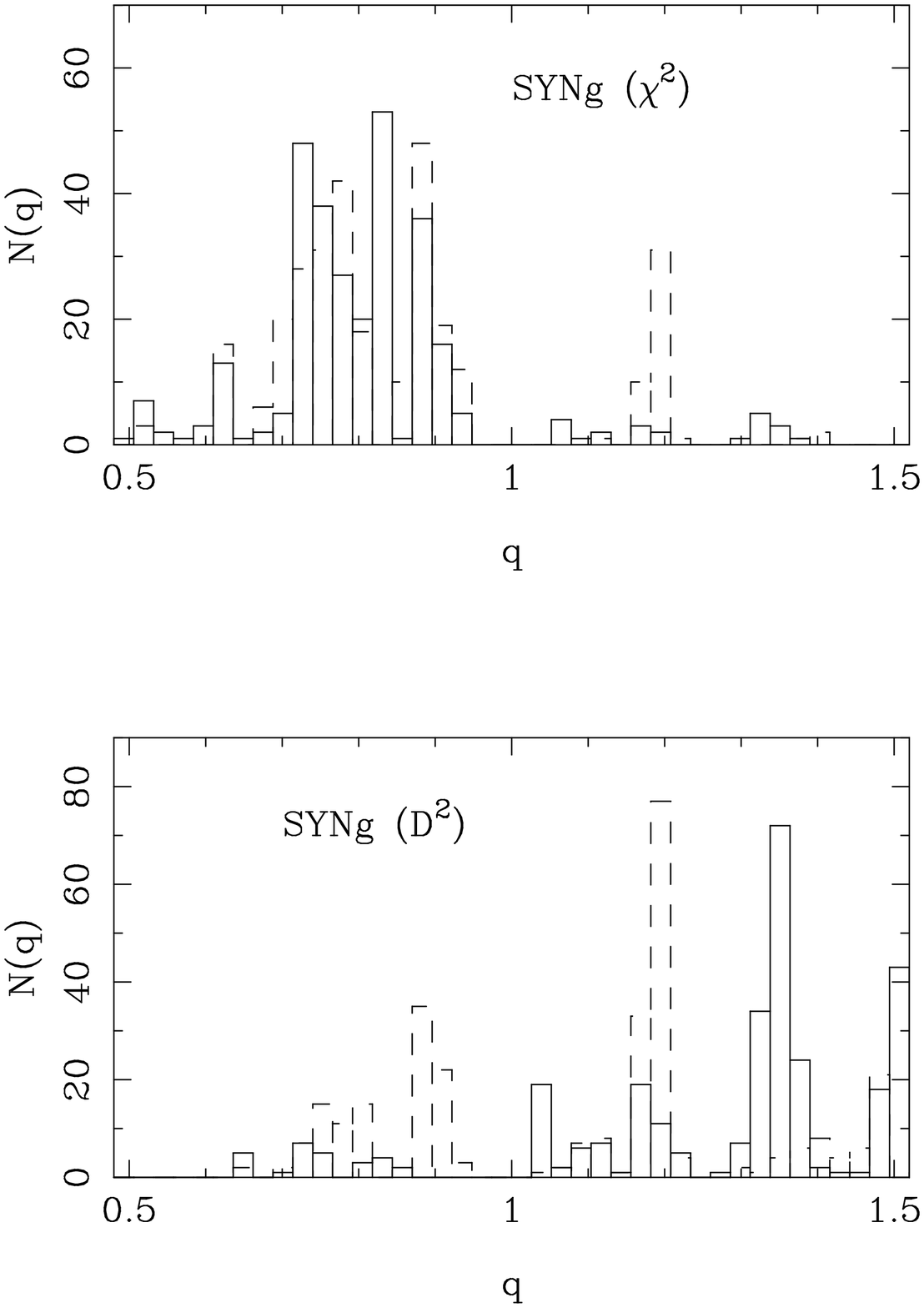}}
\caption[]{NORMAL (dashed lines) and BOOTSTRAP (solid lines) histograms associated with 
repetitions of the SYNg experiment.}
\end{figure}

\subsection{Results}

Using the experiments in Table 1 together with the scheme in section 3.1 and Appendix A, it 
might be analyzed the feasibility of an accurate estimate of the source size ratio. We consider 
a large rectangle in the ($t_0$,$q$) plane: 1450 $\leq t_0 \leq$ 1550 and 0.5 $\leq q \leq$ 
1.5, which includes the true values of $t_0$ (1483) and $q$ (0.8). 

\subsubsection{Light curves similar to the GLITP observations}

As we have about 50 $VR$ data in a period of 125 days (4 months), the typical separation 
between adjacent dates is of 2.5 days. Therefore, 2--3 days seems a good range for both the bin 
semiwidth ($\alpha$) and the decorrelation length ($\delta$). From the SYNa data set, taking 
$\alpha$ = 2.5 days, the $\chi^2$ minimization leads to a best value of $q$ = 1.5 ($t_0$ = 
1498), whereas from NORMAL and BOOTSTRAP repetitions and the minimum $\chi^2$ method, we derive
the distributions of $q$ values that appears in Figure 6 (top panel). In order to make the 
BOOTSTRAP repetitions, we use a 3--point filter. The NORMAL (dashed lines) and BOOTSTRAP (solid
lines) distributions are consistent each other. There are dominant peaks at the $q$ = 1.5 edge,
secondary features around $q$ = 0.9 and zero probabilities along the $q <$ 0.85 interval. The 
ratios $q >$ 1 are mainly derived from a relatively small number of $RV$ pairs (in general, $N 
<$ 40, and sometimes, $N \sim$ 25). Another alternative technique is the minimum dispersion 
method. When it is applied to the SYNa data (using $\delta$ = 2.5 days), the best value is 
$q$ = 1.5 ($t_0$ = 1523). The new best solution for $q$ is equal to the $\chi^2$ best solution 
for that relevant parameter, and both of them are far from the true value. From NORMAL and 
BOOTSTRAP repetitions (using a 3--point filter) and $D^2$ minimization, new distributions are 
inferred and showed in Fig. 6 (bottom panel). The new NORMAL (dashed lines) and BOOTSTRAP 
(solid lines) histograms are very enhanced at the edges. Now, there is not any structure close 
to $q$ = 0.8. Most the ratios $q \sim$ 0.5 are associated with negative amplifications ($b <$ 
0), and some ratios $q \sim$ 1.5 too. The $D^2$ results are even worse than the results from 
the $\chi^2$ minimization. 

Through the SYNb data, the $\chi^2$ best solution is $q$ = 0.98 ($t_0$ = 1456), and the 
results from the $\chi^2$ minimization and the repetitions (NORMAL and BOOTSTRAP using a 
3--point filter) are presented in Figure 7 (top panel). In all figures, dashed lines trace 
NORMAL distributions and solid lines describe BOOTSTRAP results. Regarding the distributions in
the top panel of Fig. 7, the new histograms are more homogeneous. However, there is zero 
probability at $q <$ 0.85, and the ratios $q >$ 1 are usually inferred from $N <$ 40 (in some 
cases, $N \sim$ 25). Sometimes, we simultaneously obtain a value of q larger than one and a 
negative amplification. On the other hand, the $D^2$ best solution is $q$ = 1.5 ($t_0$ = 1536).
From the minimum dispersion method, we deduce the histograms in Fig. 7 (bottom panel). We 
remark that there are no repetitions leading to the value $q$ = 0.8 (true ratio), and some 
extreme values of $q$ ($q \sim$ 0.5 or 1.5) are related to negative amplifications.

Using the SYNc data and $\alpha$ = $\delta$ = 2.5 days, the $\chi^2$ and $D^2$ best solutions
are $q$ = 0.85 ($t_0$ = 1535) and $q$ = 0.52 ($t_0$ = 1548), respectively. The NORMAL and 
BOOTSTRAP (3--point filter) histograms are plotted in Figure 8. We note that the results from 
the SYNc light curves are not very different to the results by means of the SYNa and SYNb
data sets. To sum up, with the three experiments (associated with three hypothetical 
observatories), the distributions from the minimum $\chi^2$ technique are not bell--shaped and 
centered on a ratio near the true value (top panels of Figs. 6--8). Instead of that good 
behaviour, we obtain rare distributions, which are characterized by the absence of ratios at 
$q <$ 0.85 and the presence of artifacts in the range 0.85--1.5. The false signals in the 0.85 
$\leq q \leq$ 1.5 interval can be distributed in different ways, but curious structures around 
$q$ = 0.9 are always present. These artifacts are secondary features in the top panel of Fig. 6,
prominent features in the top panel of Fig. 7 and dominant structures in Fig. 8 (top panel). In
the histograms from the $D^2$ minimization (bottom panels of Figs. 6--8), there are dominant 
peaks at the edges and negligible signals around the true value ($q$ = 0.8). Here as in other
parts of the paper, we do not show the results from the $\epsilon^2$ minimization (see sections
3.1 and A.3), because the minimum $\epsilon^2$  method works better than the $D^2$ 
minimization, but a little worse than the minimum $\chi^2$ method. As a global conclusion, 
using our framework and GLITP--like data sets, we cannot measure the visible--to--red ratio.

Although the simulations consistent with the GLITP observations indicate the non-viability of
a measurement of $q$, we compare the best values and distributions from the simulations and the
results from the GLITP data. Using the GLITP brightness records, we find that the $\chi^2$ 
($\alpha$ = 2.5 days) best solution is $q$ = 0.91 ($t_0$ = 1546). We also derive the $q$ 
distributions ($\chi^2$) in Figure 9 (top panel). The ratios $q >$ 1 are usually derived 
through $<$ 40 $RV$ pairs, and in some cases, only $\sim$ 25 $RV$ pairs are used. A slight 
correlation between large ratios ($q >$ 1) and negative amplifications ($b <$ 0) is another 
property of the results from the NORMAL and BOOTSTRAP (3--point filter) repetitions. There are 
no signals at $q <$ 0.85. However, the 0.85--1.5 range includes extended signals and dominant 
features around $q$ = 0.9. Apart from the $\chi^2$ minimization, the $D^2$ ($\delta$ = 2.5 days)
best solution is $q$ = 0.5 ($t_0$ = 1540, $b <$ 0). In Fig. 9 (bottom panel), we show the 
corresponding $q$ histograms. The ratios $q \sim$ 0.5 are mainly associated with $b <$ 0, and 
sometimes, $b < 0$ for $q >$ 1. Moreover, the distributions are enhanced at the edges. Taking 
into account the knowledge from the GLITP--like simulations (see here above), all the 
structures in Fig. 9 could be false features. Even the dominant features in the signals from 
the minimum $\chi^2$ method may be due to the limitations of the framework for the GLITP data 
set.  

\subsubsection{Sampling properties}

From the SYNd data set (see Table 1) and the techniques, we try to measure the ratio $q$.
Using $\alpha$ = 2.5 days, the $\chi^2$ best solution is $q$ = 1.24 ($t_0$ = 1481). In addition
to this best value, the $q$ distributions from the $\chi^2$ minimization and 300 repetitions 
(NORMAL and BOOTSTRAP) are depicted in Figure 10 (top panel). In the BOOTSTRAP procedure, as 
usual, it is used a 3--point filter. The histograms in the top panel of Fig. 10 are clearly 
different to the distributions in the top panels of Figs. 6--8. In the new signals, we see 
dominant structures at $q <$ 0.7 and do not see the artifacts around $q$ = 0.9, which suggests 
the existence of a relation between the gaps in the GLITP--like curves and the trends in the 
top panels of Figs. 6--8. In any case, the fraction of ratios in the 0.7--0.9 interval (i.e., 
the probability that the true value will fall within the 0.7 $\leq q \leq$ 0.9 range) is very 
small. Therefore, as $P$(0.7 $\leq q \leq$ 0.9) $\leq$ 10\% and the true value is $q$ = 0.8, we
basically obtain false signals. From the minimum dispersion method, we infer a best ratio of 
1.5 ($t_0$ = 1525) and deduce the histograms in the bottom panel of Fig. 10. There are dominant
peaks at the $q$ = 1.5 edge and very small probabilities at $q <$ 1.2. We remark that the 
improvement in the sampling homogeneity is not sufficient, since the new best values and $q$ 
distributions do not permit to estimate the visible--to--red ratio. 

By means of either a large collaboration including several observatories around the world or a
space telescope, in each optical band, we can get a quasi--continuous sampling of one frame per 
day. This "ideal" sampling is assumed in the SYNe experiment. The time resolution is roughly 
increased by a factor of 2, so 1.2 days is a reasonable choice for both the bin semiwidth 
($\alpha$) and the decorrelation length ($\delta$). The $\chi^2$ minimization leads to a very 
promising best value for both the ratio and the time of caustic crossing: $q$ = 0.79 and $t_0$ 
= 1484. However, the distributions of $q$ values have not a behaviour as good as expected. In 
Figure 11 (top panel), the NORMAL and BOOTSTRAP histograms ($\chi^2$) appear. Around $q$ = 0.8,
there are prominent peaks with relatively small probabilities of $P$(0.7 $\leq q \leq$ 0.9) 
$\approx$ 23--24\%. These significant features are not dominant structures, but structures 
surrounded by other similar features. As a result of the existence of several similar features,
a fair measurement of $q$ cannot be attained. When the minimum $D^2$ method is applied to the 
SYNe data, the best value is $q$ = 1.48 ($t_0$ = 1510). From NORMAL and BOOTSTRAP repetitions
and $D^2$ minimization, we infer disappointing distributions of $q$ (see the bottom panel of 
Fig. 11). It becomes apparent the absence of a significant feature in the surroundings of $q$ =
0.8 (true value), and of course, the histograms ($D^2$) are strongly biased. Finally, we 
conclude that our framework does not work in a proper way, even with a substantial improvement 
in the sampling properties. 

\subsubsection{Flux errors}

In the SYNf experiment, the flux errors are lowered in a factor 2 (see Table 1 and Fig. 4). 
From the $\chi^2$ minimization, taking $\alpha$ = 2.5
days, the best ratio is $q$ = 0.73 ($t_0$ = 1489). Using the minimum $\chi^2$ technique with 
the usual time resolution ($\alpha$ = 2.5 days and 3--point filter in BOOTSTRAP repetitions), 
we obtain two $q$ distributions that are depicted in Figure 12 (top panel). In the top panel of
Fig. 12, we see encouraging BOOTSTRAP results. From the BOOTSTRAP repetitions, a dominant peak 
around $q$ = 0.76 appears. This central value (0.76) is in good agreement with the best ratio 
(0.73), and moreover, the true ratio (0.8) is included in the $q$ range for the dominant 
feature. The NORMAL results are worse than the BOOTSTRAP results, because the NORMAL main spike 
is placed at the $q$ = 0.5 edge. Using the BOOTSTRAP histogram, we can derive the first 
estimate of the source size ratio. Following the procedure that is described in section 3.1, 
$q$ = 0.76 $\pm$ 0.05. To obtain the measurement, we ignore the signal at $q <$ 0.6 and $q >$ 
0.9. However, unfortunately, the possibility of $\sim$ 3--10\% measurements of the source size 
ratio is not supported by other experiments similar to the SYNf one. For example, from ten 
new synthetic experiments and the $\chi^2$ minimization, only three best ratios are included in
the promising 0.65 $\leq q \leq$ 0.85 interval. In five cases, the best ratio is of about 0.5, 
whereas in two cases, the best value is close to 1.4. Therefore, if we consider ten 
hypothetical observers measuring the source size ratio each of them (via $\chi^2$/BOOTSTRAP), 
several best estimates and distributions of $q$ will be in serious disagreement with the true 
ratio. Although some particular observers are successful, most observers fail in the 
determination of $q$. Through the SYNf--like experiments, we also derive $\chi^2_{min}$ values 
in the interval 0.25--0.60. These results suggest that we deal with seven biased "superfits", 
which are not related to the true physical scenario. The {\it observational} noise and 
discontinuous sampling are the cause of the superb correlations between the $V$ fluxes and the 
$R$ ones.  
From the minimum dispersion method, taking $\delta$ = 2.5 days, the best 
ratio is $q$ = 1.15 ($t_0$ = 1523). The NORMAL and BOOTSTRAP repetitions lead to poor 
histograms. These $D^2$ histograms appear in Fig. 12 (bottom panel). In spite of the small peak
in the surroundings of the true value, it is apparent that the distributions are biased. From 
the NORMAL and BOOTSTRAP distributions of $q$ ($D^2$), a false ratio exceeding the critical 
value ($q$ = 1) is strongly favoured. If we compare the results in Fig. 12 and the 
distributions in Fig. 10 (from a data set with similar sampling and larger errors), it is clear 
that the decrease of the flux errors leads to a global improvement. With smaller errors, we 
find stronger signals in the proximity of the true ratio.

As the {\it photometric} errors seem to have a significant influence on the $q$ distributions, 
finally, we explore the ability of the methodology with extremely accurate {\it photometric} 
data. For this final effort, the SYNg data set is a suitable tool. For the SYNg data, the
$\chi^2$ minimization gives a best solution: $q$ = 0.79 ($t_0$ = 1484), while for the NORMAL 
and BOOTSTRAP repetitions, the technique also works well. In Figure 13 (top panel), the 
corresponding distributions are plotted.
We see dominant structures around $q$ = 0.80 (NORMAL and BOOTSTRAP central value). With the 
$q$ resolution in the top panel of Fig. 13, the main features have not a nice shape, but they 
contain most the best solutions. Our $\chi^2$/NORMAL\&BOOTSTRAP measurement is of $q$ = 0.80 
$\pm$ 0.08 (ignoring the signal at $q <$ 0.65 and $q >$ 0.95). From ten {\it monitorings} 
similar to the SYNg experiment and the minimum $\chi^2$ method, we get convincing 
results: nine best ratios are within the 0.70 $\leq q \leq$ 0.92 interval, and only one best 
ratio has a biased value of $q$ = 0.52. Even in this last case ($q$ = 0.52), the NORMAL and 
BOOTSTRAP histograms have relatively good behaviours. The NORMAL distribution shows a dominant 
spike close to 0.5, which contains about 50\% of the best ratios. However, the rest of ratios 
(about 50\%) are mainly placed in the 0.65 $\leq q \leq$ 0.95 range. In the BOOTSTRAP 
distribution, there is also an extended feature within the 0.65 $\leq q \leq$ 0.95 range, which
includes about 70\% of the ratios. The highest spike at $q \sim$ 0.5 only contains about 30\% 
of the ratios. In other words, the hypothetical observer would find doubtful results, since 
there are evidences for two different values: $q \sim$ 0.5 (artifact) and $q \sim$ 0.8 (true 
ratio). Apart from this troublesome situation, any possible observer has a very high 
probability (about ninety per cent) of measuring a fair and accurate visible--to--red ratio 
($\sim$ 10\% measurement). For the nine SYNg--like experiments leading to non--biased values
of $q$, the $\chi^2_{min}$ varies from 0.95--1.20, i.e., we infer a reasonable $\chi^2_{min}$ 
interval and all is ok. However, $\chi^2_{min}$ = 0.6 from the SYNg--like experiment associated
with a strange value of $q$. The strange ratio is related to a very small $\chi^2$ 
("superfit"), and both ($\chi^2_{min}$ and $q$) are due to the noise and sampling.
From the minimum $D^2$, the best ratio is $q$ = 0.80 ($t_0$ = 1484).
This is the only case in which both the $\chi^2$ and $D^2$ best estimates are nearby each other 
and the true value. New $D^2$ histograms appear in Fig. 13 (bottom panel). The two 
distributions in the bottom panel of Fig. 13 are really rare. We see the true signal inside the
0.65 $\leq q \leq$ 0.95 interval and other features at $q >$ 1. The main structures are related
to ratios in the $\sim$ 1.1--1.4 range. Why are the $D^2$ histograms so rare?. From the 10 
synthetic data sets similar to the SYNg one and the minimum $D^2$ method, we infer 
surprising results: two ratios exceeding the critical value (i.e., larger than 1) and eight 
ratios larger than 0.74 and smaller than 0.91. This independent $q$ distribution (based on real
repetitions, i.e., using the true underlaying signal) indicates that the NORMAL and BOOTSTRAP 
procedures may be unsuitable with extremely accurate light curves and the $D^2$ minimization. 
Finally, we note that future monitoring projects from modern ground--based or space telescopes 
can lead to $\sim$ 10\% measurements of the visible--to--red ratio. To be successful in the 
accurate determination of $q$, a reasonable sampling and a few $\mu$Jy uncertainties are 
required.

\section{Summary and discussion}

We present a new framework to analyze the structure of the optical compact source of a lensed 
QSO. When a microlensing high--magnification event (HME) is produced in one of the QSO 
components, assuming that the compact emission regions have different sizes in different 
wavelengths, the multiband light curves of the HME can be used to measure the source size 
ratios (e.g., Wambsganss \& Paczy\'nski 1991). In this paper, we deal with a kind of HMEs: the 
special high--magnification events (SHMEs). This family includes the well--known caustic 
crossing as well as other situations, e.g., the two--dimensional maximum crossing. Our method 
has the advantage that finds the source size ratios in a direct and model--independent 
(stationary source model) way and without complex computation procedures. From the brightness 
records of a caustic crossing event (CCE), the deconvolution technique leads to a richer 
information, because the method enables  to retrieve the one--dimensional intrinsic 
luminosity profiles (e.g., Grieger, Kayser \& Schramm 1991). However, the determination of the 
1D intrinsic luminosity profiles is not a fair and simple task, and the problem is related to 
complex inversion procedures. To infer a source size ratio, we propose a straightforward 
cross--correlation between the records in the two optical bands, so our procedure has some 
resemblance to the classical time delay measurement. In order to measure the visible--to--red 
ratio ($q = R_V/R_R$), we also introduce several suitable tools, which can be applied to derive
another ratio ($R_U/R_B$, $R_B/R_V$, ...). 

The power of the new scheme is tested from synthetic light curves that are related to the
$V$--band and $R$--band GLITP microlensing peaks in the flux of Q2237+0305A (Alcalde et al. 
2002). Very recently, assuming that the GLITP/Q2237+0305A fluctuations are due to a CCE, 
Shalyapin et al. (2002) and Goicoechea et al. (2003) analyzed the nature and size of the 
optical compact source, as well as the central mass and accretion rate associated with the 
favoured model (standard accretion disk). In this work, the GLITP/Q2237+0305A records are also 
associated with a CCE (for a discussion on the origin of the GLITP/Q2237+0305A data set, see
here below). To generate synthetic datasets, we take underlying signals in agreement with the 
GLITP observations (reduced $\chi^2$ values close to 1). They correspond to $p$ = 3/2 
power--law source profiles crossing a fold caustic, so the source size ratio is taken as $q$ = 0.8 
(see Tables 1--3 in Shalyapin et al. 2002). Once an underlying signal is made, we add {\it 
observational} random noise, which is characterized by a normal standard deviation. This random
noise must incorporate the pure {\it observational} uncertainty and the day--to--day intrinsic 
variability. We remark that the scheme is based on a stationary source model, and thus, the 
underlying signal cannot include any intrinsic variation. While in some synthetic experiments, 
the flux errors and sampling properties agree with the GLITP photometric uncertainties and 
sampling, in other experiments, the influence of the sampling properties and flux errors is 
studied in detail. We find that GLITP--like datasets are not suitable for measuring the 
visible--to--red ratio. Even with a dramatic improvement in the sampling, our framework does 
not lead to convincing results. However, if the flux uncertainties are significantly lowered, 
the scheme works in an accurate way. From $VR$ light curves with a few $\mu$Jy uncertainties, 
we can infer $\sim$ 10\% measurements of $q$. Assuming the NOT uncertainties for Q2237+0305A 
(of about 10 $\mu$Jy) as mainly due to pure observational noise, it would be viable to achieve 
smaller errors using the current superb--telescopes (the best ground--based telescopes or the 
Hubble Space Telescope). Therefore, there are no technological obstacles to get accurate 
estimates of the visible--to--red ratio for QSO 2237+0305. The possible presence of very rapid 
intrinsic variability with relatively large amplitude would be the only serious obstacle. Using the
$\chi^2$ minimization, one can obtain an accurate value of $q$ in two ways: either from only one 
monitoring and standard techniques to infer the error in $q$, or from the best solutions 
corresponding to several datasets of different observatories. However, using the minimum 
dispersion method, the standard repetitions of an individual experiment do not seem to lead to 
good results, and one must focus on the best solutions from different experiments. In general, the
$\chi^2$ minimization works better than the minimum dispersion method, and the $\epsilon^2$ 
minimization is a technique with intermediate quality.

Sampling and flux errors aside, other factors may determine the ability of the methodology.
For example, the time coverage of the SHME. We only test microlensing peaks lasting $\sim$ 
100 days, but longer and larger variations could lead to an accurate determination of the 
ratio, without need for improving the uncertainties. Nevertheless, it is hard to imagine a 
long period of about 1 year in which the source QSO does not vary. From a long--timescale
monitoring, we would observe a dirty SHME, i.e., true microlensing fluctuations that are 
contaminated by some intrinsic variation. Moreover, it may be difficult to detect a pure SHME, 
since a long--timescale event may include variability from either several features in the 
magnification pattern or random stellar motions in the deflector. Another possibility is the
determination of $q$ from very fast microlensing events. For a given level of noise, some probes 
indicate that the fastest underlaying signals enable the best measurements of the ratio. Therefore, 
faster events in  Q2237+0305A as well as very fast events in another component of that system or 
other lensed QSO would represent more favourable situations.

Before to reliably apply the scheme, a key point is to confirm that the observed records are 
very probably related to a clean and pure SHME. This task is not so easy, and currently, even 
the origin of the GLITP/Q2237+0305A fluctuations is not a totally clear matter. The global flat
shape for the $VR$ GLITP light curves of Q2237+0305D (the faintest component of the system) 
indicates the absence of a global intrinsic variation. Therefore, the GLITP/Q2237+0305A peaks 
seem to be clean microlensing fluctuations. On the other hand, the $V$--band and $R$--band 
GLITP light curves of Q2237+0305A trace the regions around the maxima of the $VR$ fluctuations.
As the peaks are highly asymmetric and correspond to a prominent event (observations by the OGLE 
team), they were associated with a CCE (a kind of SHME) since were discovered. The CCE hypothesis 
led to very reasonable results for the source structure (Shalyapin et al. 2002; Goicoechea et al. 
2003) , which are an {\it a posteriori} support for the initial hypothesis.  But do they really 
correspond to a pure CCE?. Kochanek (2004) studied the source trajectories that agree with the
whole $V$--band OGLE light curves for Q2237+0305A--D. His results for the origin of the 
prominent event in the A component are a bit disappointing, because the best trajectories in 
terms of $\chi^2$ cross over simple folds, but other relatively good trajectories pass through 
complex magnification zones (see Figs. 12--16 in Kochanek 2004). However, several issues suggest 
that the Kochanek's conclusions about the nature of the microlensing event are preliminary ones. 
First, the conclusions were based on a joint study of the four components A--D during a long period. 
Second, it was used an enlargement of the formal errors, so the pair of best trajectories on the 
magnification patterns for Q2237+0305A have excessively small values of $\chi^2$ ($\chi^2_0 = 
186-187$, $N_{dof}$ = 290), and the rest of good paths are characterized by $\Delta \chi^2 = 
\chi^2 - \chi^2_0 \geq 14-15$. In the circumstances, the use of smaller uncertainties does not seem 
unrealistic. Moreover, the new uncertainties could lead to $\Delta \chi^2 > (2N_{dof})^{1/2}$. 
Third, in order to obtain statistical conclusions, the total number of good trajectories is 
clearly small. At present, the University of Cantabria group is carrying out a deep study about
the origin of the OGLE--GLITP/Q2237+0305A event.
         
\begin{acknowledgements}
We thank R. Gil--Merino for a careful reading of the manuscript and suggestions. We also thank 
C. S. Kochanek for comments on the origin of the OGLE--GLITP/Q2237+0305A event, and F.
Almeida and F. de Sande (Depto. Estadistica, I.O. y Computacion, Universidad de La Laguna) for
valuable aid in the four--dimensional minimization of the dispersion, which was used to test the
two--dimensional one. The authors would like to thank the anonymous referee for comments on
the overall structure of the paper. This work was supported by Universidad de Cantabria funds and 
the Spanish Department for Science and Technology grant AYA2001-1647-C02. 
\end{acknowledgements}

{}

\appendix
\section{Minimization techniques}

\subsection{$\chi^2$ minimization}

The whole data set includes the {\it observed} fluxes in the $R$ band, $F_R(t_i)$, $i = 
1,2,...,N_R$, with common uncertainties $\sigma_{R}$, and the {\it observed} fluxes in the $V$ 
band, $F_V(t_j)$, $j = 1,2,...,N_V$, with {\it observational} errors $\sigma_{V}$. The 
$R$--band flux at time $t_i$, $F_R(t_i)$, is compared to the flux $a + bF_V(t'_i)$, where 
$t'_i = qt_i + (1 - q)t_0$. In general, the dilated and delayed time $t'_i$ does not coincide 
with any epoch in the $V$ band, and we estimate the value of $F_V(t'_i)$ by averaging 
the $V$--band fluxes within the bin centered on $t'_i$ with a semiwidth $\alpha$. To  
average, it is appropriate the use of weights depending on the separation between the central 
time $t'_i$ and the dates $t_j$ in the bin. For given values of $q$ and $t_0$, the number of 
possible [$F_R(t_i)$,$F_V(t'_i)$] pairs is less or equal to $N_R$, since some $V$--band bins 
may be empty. The $\chi^2$ estimator is given by 
\begin{equation}
\chi^2(a,b,t_0,q) = \frac{1}{N-4} \sum_{i=1}^{N} \frac{[F_R(t_i) - a - bF_V(t'_i)]^2}
{\sigma_{R}^2 + b^2\sigma_{Vi}^2}   ,
\end{equation}
where $N$ is the number of $RV$ pairs ($N \leq N_R$), $F_V(t'_i) = [\sum_j S_{ij}F_V(t_j)]/
\sum_j S_{ij}$, the weight--selection factors $S_{ij}$ are
\begin{equation} 
S_{ij} = \left\{
\begin{aligned}
1-\frac{|t'_i-t_j|}{\alpha} ,\;\; \rm if \it \;  |t'_i-t_j|\leq\alpha,  \\
0 ,\;\,  \rm if \it \;  |t'_i-t_j|>\alpha ,
\end{aligned}
\right.
\end{equation}
and the uncertainties in the fluxes $F_V(t'_i)$ are 
\begin{equation} 
\sigma_{Vi}^2 = \frac{\sum_j S_{ij}^2 \sigma_{V}^2}{(\sum_j S_{ij})^2}  .
\end{equation}

In principle, the $\chi^2$ estimator is a function of four parameters $(a,b,t_0,q)$. However, 
as the parameter $a$ is entered in Eq. (A.1) in a simple way, it is possible to obtain an
analytical constraint $a = a(b,t_0,q)$ from the minimization condition $\partial \chi^2/
\partial a = 0$. One finds $a = P - bQ$, where
\begin{equation} 
P = \left[\sum_{i=1}^{N} \frac{F_{Ri}}{\sigma_{R}^2 + b^2\sigma_{Vi}^2} \right]/
\left[\sum_{i=1}^{N} \frac{1}{\sigma_{R}^2 + b^2\sigma_{Vi}^2} \right]  ,
\end{equation}
and
\begin{equation} 
Q = \left[\sum_{i=1}^{N} \frac{F_{Vi}}{\sigma_{R}^2 + b^2\sigma_{Vi}^2} \right]/
\left[\sum_{i=1}^{N} \frac{1}{\sigma_{R}^2 + b^2\sigma_{Vi}^2} \right]  .
\end{equation}
Thus, we search for the minimum of $\chi^2$ in a 3D parameter space, i.e., we minimize the
function $\chi^2 = \chi^2[a(b,t_0,q),b,t_0,q]$.

\subsection{Minimum dispersion method}

Our $R$--band data are modelled as $F_{Ri} = s(t_i)
+ \epsilon_R(t_i)$. Here, $s$ and $\epsilon_R$ denote the true $R$--band signal and the unknown
$R$--band errors, respectively. In a consistent way (see Eqs. 6 and 8), the $V$--band data 
should be modelled as $F_{Vj} = \{s[t_j/q + (1 - 1/q)t_0] - a\}/b + \epsilon_V(t_j)$. These two
series are combined into one for every fixed value of an offset $a$, an amplification $b$, a 
characteristic time $t_0$ and a dilation factor $q$. In the combined serie, $N_R$ data are the 
$F_{Ri}$ values at times $t_i$, whereas the rest of data ($N_V$) are the $a + bF_{Vj}$ values 
at dates $t'_j = t_j/q + (1 - 1/q)t_0$. Each combined curve includes $N_R + N_V$ fluxes and 
times. The dispersion of the combined curve is
\begin{equation} 
D^2(a,b,t_0,q) = \frac{\sum_{i=1}^{N_R}\sum_{j=1}^{N_V} S_{ij} W_{ij} (F_{Ri} - a - 
bF_{Vj})^2}{\sum_{i=1}^{N_R}\sum_{j=1}^{N_V} S_{ij} W_{ij}}  ,
\end{equation}
where $S_{ij}$ are weight--selection factors defined by
\begin{equation} 
S_{ij} = \left\{
\begin{aligned}
1-\frac{|t_i-t'_j|}{\delta} ,\;\; \rm if \it \;  |t_i-t'_j|\leq\delta,  \\
0 ,\;\,  \rm if \it \;  |t_i-t'_j|>\delta ,
\end{aligned}
\right.
\end{equation}
and $W_{ij} = 1/(\sigma_{R}^2 + b^2\sigma_{V}^2)$ are the statistical weights. We note that
all the ($i$,$j$) pairs have equal statistical weight, and in this special case, the $W_{ij}$ 
factors do not play a role in the $D^2$ estimator. The main difference between the old problem 
(estimation of the best time delay) and the new one (estimation of the best source size ratio)
lies in the dilation factor that is absent in delay studies. 
 
In the minimization process, one can also reduce the dimension of the parameter space (see the 
end of section A.1). We search for the minimum dispersion in a 2D parameter space, since 
there are analytical constraints $a = a(t_0,q)$ and $b = b(t_0,q)$. These constraints are 
inferred from the system of equations: $\partial D^2/\partial a = \partial D^2/\partial b = 0$.
In a straightforward way, the system leads to $a = P - bQ$ and $b = (X - PQ)/(S - Q^2)$, being
\begin{equation}
P = \sum_i\sum_j S_{ij} F_{Ri} / \sum_i\sum_j S_{ij}  ,
\end{equation}
\begin{equation}
Q = \sum_i\sum_j S_{ij} F_{Vj} / \sum_i\sum_j S_{ij}  ,
\end{equation}
\begin{equation}
S = \sum_i\sum_j S_{ij} F_{Vj}^2 / \sum_i\sum_j S_{ij}  ,
\end{equation}
\begin{equation}
X = \sum_i\sum_j S_{ij} F_{Ri} F_{Vj} / \sum_i\sum_j S_{ij}  .
\end{equation}

\subsection{Minimum modified dispersion method}

We also propose a modified dispersion. The basic difference lies in the fact that we do not use
the usual terms $(F_{Ri} - a - bF_{Vj})^2$, but the normalized ones 
$(F_{Ri} - a - bF_{Vj})^2/(\sigma_{R}^2 + b^2\sigma_{V}^2)$. The new estimator has an 
expression 
\begin{equation} 
\epsilon^2(a,b,t_0,q) = \frac{\sum_{i=1}^{N_R}\sum_{j=1}^{N_V} S_{ij} W_{ij} (F_{Ri} - a - 
bF_{Vj})^2}{\sum_{i=1}^{N_R}\sum_{j=1}^{N_V} S_{ij}}  .
\end{equation}
The $\epsilon^2$ estimator depends on four parameters: $a$, $b$, $t_0$ and $q$, 
but we can use some constraints and simplify the 4D minimization process. From  $\partial 
\epsilon^2/\partial a = 0$, it is inferred the relationship
\begin{equation} 
a = \frac{\sum_i\sum_j S_{ij} (F_{Ri} - bF_{Vj})}{\sum_i\sum_j S_{ij}}  .
\end{equation}
If we denote the {\it averages} of the light curves as
\begin{equation}
P = \sum_i\sum_j S_{ij}  F_{Ri} / \sum_i\sum_j S_{ij} ,\ \
Q = \sum_i\sum_j S_{ij}  F_{Vj} / \sum_i\sum_j S_{ij} ,
\end{equation}
then the expression for the parameter $a$ takes the simple appearance
\begin{equation}
a = P - bQ .
\end{equation}
Using  $\partial \epsilon^2/\partial b = 0$, it is derived a second interesting constraint. The
new constraint can be written as 
\begin{equation}
\sum_i\sum_j S_{ij}  (F_{Ri} - a - bF_{Vj}) \left( b\sigma_V^2 F_{Ri} + 
\sigma_R^2 F_{Vj} \right) = 0  .
\end{equation}
In order to simplify the expression (A.16), we introduce the deviations of the fluxes from the 
{\it averages}. Thus, 
\begin{equation}
\delta F_{Ri} = F_{Ri} - P, \ \ \delta F_{Vj} = F_{Vj} - Q .
\end{equation}
From Eqs. (A.15), (A.16) and (A.17), we finally obtain
\begin{equation} 
\begin{split}
- b^2\sigma_V^2 \sum_i \sum_j S_{ij}  F_{Ri} \delta F_{Vj} + b \left( \sigma_V^2 \sum_i 
\sum_j S_{ij}  F_{Ri} \delta F_{Ri} -  \sigma_R^2 \sum_i \sum_j S_{ij}  F_{Vj} \delta F_{Vj} 
\right) + 
\\ \sigma_R^2 \sum_i \sum_j S_{ij} F_{Vj} \delta F_{Ri} = 0  .
\end{split}
\end{equation}
Now it is clear that one can work in a 2D parameter space. For a given pair ($t_0$,$q$), 
through Eqs. (A.15) and (A.18), we can straightway derive the solutions ($a$,$b$) that minimize 
the $\epsilon^2$ estimator. 

\end{document}